\documentclass[a4paper,12pt]{article}
\usepackage{amsmath,amsfonts,amssymb,amsthm,amstext,amscd}
\usepackage{latexsym}
\usepackage{hyperref}
\usepackage{graphicx}
\usepackage{caption}
\usepackage{comment}
\usepackage{graphicx}
\usepackage{cite}
\usepackage{color}
\usepackage{setspace}
\usepackage[mathscr]{eucal}
\usepackage{xcolor}
\usepackage{titletoc}
\usepackage{graphicx,txfonts}

\newcommand{\heart}{\ensuremath\varheartsuit}

\newcommand{\mrd}{\mathrm{d}}

\newcommand{\mr}[1]{\mathrm{#1}}
\renewcommand{\H}{{_\mathrm{H}}}

\definecolor{darkred}{rgb}{0.9,0.05,0.05}
\definecolor{darkblue}{rgb}{0.05,0.05,0.6}
\definecolor{darkgreen}{rgb}{0.05,0.6,0.05}
\definecolor{brightgreen}{rgb}{0.1,0.9,0.1}

\renewcommand*{\eqref}[1]{%
	\begingroup
	\hypersetup{
		linkcolor=darkblue,
		linkbordercolor=darkblue,
	}%
	\textcolor{darkblue}{(\ref{#1})}%
	\endgroup
}
\hypersetup{linkcolor=black,citecolor=darkgreen,urlcolor=black,colorlinks=true}

\begin{document}

\setlength{\skip\footins}{0.8cm}

\begin{titlepage}

\vspace{0.5cm}

\newcommand\blfootnote[1]{%
	\begingroup
	\renewcommand\thefootnote{}\footnote{#1}%
	\addtocounter{footnote}{-1}%
	\endgroup
}

\begin{center}
\renewcommand{\baselinestretch}{1.5}  
\setstretch{1.5}

{\fontsize{17pt}{12pt}\bf{A Universal Smarr Formula via Coupling Constants}}
 
\vspace{9mm}
\renewcommand{\baselinestretch}{1}  
\setstretch{1}

\centerline{  {Kamal Hajian$^{\ast\dagger}$\footnote{kamal.hajian@uni-oldenburg.de}},  {Bayram  Tekin$^{\heart}$\footnote{bayram.tekin@bilkent.edu.tr}}, and {Onur Uçanok$^{\ast}$\footnote{ucanoko@metu.edu.tr}} }
\vspace{4mm}
\normalsize
$^\ast$\textit{Department of Physics, Middle East Technical University, 06800, Ankara, Turkey}\\
$^\dagger$\textit{Institute of Physics, University of Oldenburg, P.O.Box 2503, D-26111 Oldenburg, Germany}\\
$^\heart$\textit{Department of Physics, Bilkent University, 06800, Ankara, Turkey}\\
\vspace{5mm}
\begin{abstract}\noindent
In gravitational theories containing matter fields and higher-derivative corrections, the standard Smarr formula often fails unless all dimensionful couplings are incorporated consistently. Traditionally, parameters such as the cosmological constant or the coefficients of higher-derivative terms are regarded as immutable features of the theory and therefore excluded from the thermodynamic phase space. In our recent work, we developed a fully general framework that promotes every such coupling to a dynamical, freely varying parameter of black hole solutions. This is accomplished by introducing, for each coupling, an auxiliary scalar and gauge field, through which the coupling appears as a conserved charge associated with the global sector of an emergent gauge symmetry. The corresponding conjugate variables naturally arise as electric potentials evaluated at the black hole horizon. As a result, the first law and the Smarr relation acquire additional, systematically determined contributions, yielding a consistent and universal extension of black hole thermodynamics. We illustrate the validity of this construction by revisiting several black hole examples in the literature where the Smarr formula remains inconsistent even after treating the cosmological constant as a thermodynamic variable. Our analysis shows that only by including all dimensionful couplings in this generalized manner can one obtain an internally consistent Smarr relation, thereby providing the foundation for a truly universal formulation of black hole thermodynamics.
\end{abstract}

\end{center}

\let\newpage\relax

\end{titlepage}

\section{Introduction}

Black hole thermodynamics within General Relativity \cite{Bekenstein:1973ft,Hawking:1976rt,Bardeen:1973gd}, in addition to the four laws, also includes the celebrated Smarr formula \cite{Smarr:1972kt}, an integrated relation between the thermodynamic coordinates and their conjugates, unlike the differential first law. It is consistent with the first law but does not directly follow from it; one also needs dimensional scaling, that is, the homogeneity of the black hole's mass with respect to other conserved charges. This critical relation is often invalid in generic theories of gravity that include matter fields, even if the first law holds, particularly when contributions from the theory's coupling constants are not adequately accounted for. These coupling constants, such as the cosmological constant $\Lambda$ \cite{Einstein:1917} or other dimensionful parameters present in the Lagrangian, are inherently fixed parameters (at least at the classical level) defining the theory and are traditionally not subject to variation. This state of affairs needs to be understood: is the Smarr formula specific to General Relativity, or is it a universal relation among the thermodynamic coordinates and their conjugates? Here, we argue that it is the latter, after properly enlarging the theory to include other dimensionful coupling constants beyond Newton's constant. 

A known difficulty arises with the cosmological constant $\Lambda$. Although it appears as a fixed parameter in the Lagrangian, it behaves like a thermodynamic variable that is typically tied to the properties of a solution. To preserve the consistency of the Smarr relation, variations of $\Lambda$ must therefore be included in the first law of black hole thermodynamics \cite{Henneaux:1984ji,Henneaux:1989zc,Teitelboim:1985dp,Brown:1986nw}. This perspective led, in the early 2010s, to the development of black hole chemistry, where $\Lambda$ is reinterpreted as pressure, contributing through a $V\delta P$ term in the first law. The term ``chemistry" is used because, with this identification, the black hole mass must be treated as the enthalpy rather than the internal energy—the quantity minimized in chemical systems (see the excellent reviews \cite{Mann_Kubiznak, Mann:2025xrb} and references therein). 

Treating the cosmological constant as a pressure introduces further challenges to our understanding of black hole thermodynamics. The rationale for varying a fundamental constant of the theory is unclear. Besides, its conjugate chemical potential lacks a robust geometric definition; unlike the surface area of the event horizon, the black hole's geometric volume is not an invariant quantity. Moreover, while the black hole mass represents the conserved charge associated with time translation symmetry, its enthalpy is not regarded as a conserved quantity in thermodynamics. The situation becomes more intricate once one recognizes that, to preserve the Smarr formula, not only must the cosmological constant be included, but also all dimensionful couplings. At the same time, the pressure can account for only one of them.  In addition, the quantity $V$, as the conjugate potential of $\delta P$, is proposed to be fundamentally different from the other terms appearing in the first law: (i) it is an extensive quantity, whereas other potentials, such as temperature and angular velocity, are intensive; (ii) by definition, it is not a property of the horizon itself but of the region enclosed by it (otherwise, it loses its meaning as a volume).   These issues, along with the broader problem of varying the Lagrangian's constant parameters, highlight the lack of a universal and coherent theoretical framework in the existing literature.

As a solution to this problem, a robust method was developed in  \cite{Hajian:2023bhq} that redefines the role of coupling constants, transforming them from fixed parameters of the theory into free parameters of the solutions. This approach involves modifying the Lagrangian by introducing auxiliary gauge fields, a concept previously explored for the cosmological constant in works such as \cite{Aurilia:1980xj, Duff:1980qv, Hawking:1984hk, Bousso:2000xa} and, specifically, in \cite{1, Hajian:2021hje}. By doing so, the coupling constants are elevated to conserved charges of the implemented gauge symmetry, and their conjugate chemical potentials are naturally defined as the potentials of the corresponding gauge fields on the black hole horizon, similar to the electric potential in electromagnetism.

In \cite{Hajian:2023bhq}, this method was presented as a general formulation capable of treating all coupling constants as conserved charges within black hole physics, thereby placing black hole thermodynamics on a firmer theoretical foundation. A key distinguishing feature of the proposed construction, when compared to earlier approaches, is twofold: 1) For each coupling constant (including the cosmological constant), a pair of auxiliary fields is introduced, consisting of one scalar field and one gauge field; 2) The Lagrangian associated with these auxiliary gauge fields is linear in their field strength rather than quadratic.
These distinctions are crucial for establishing a general method applicable to arbitrary coupling constants.

This systematic extension of the first law and the Smarr formula, by coupling conserved charges to their conjugate potentials, offers a novel vantage point. The work demonstrates that coupling constants are indeed conserved charges linked to the global part of the implemented gauge symmetry. The conjugate chemical potentials are defined consistently, inspired by the electric potential in Maxwell's electrodynamics. This framework allows for a comprehensive treatment in which the variation of couplings as solution parameters can be incorporated into the first law of black hole thermodynamics. The Smarr formula is derived from the extended first law using dimensional analysis and scaling arguments. 

Our purpose here is to examine this formulation in different theories and black hole solutions that have previously suffered from the absence of couplings in their Smarr formula and the first law. In Sec. \ref{Sec review}, the formulation is reviewed, and in Sec. \ref{Sec analysis}, different examples are analyzed, their corresponding new gauge fields are derived, and their thermodynamic relations are reexamined.  

\section{A review of the formulation}\label{Sec review}

Let us assume that the theory is defined by a generic $D$-dimensional action with coupling constants $\alpha_i$: $I = \int (\mathbf{L}_0 - \sum_i \alpha_i \mathbf{L}_i)$. This action is modified by introducing, for each coupling $\alpha_i$, a pair of auxiliary fields: a scalar field $\alpha_i(x)$ and a $(D-1)$-form gauge field $\boldsymbol{\mathrm{A}}_i$. The extended action $\tilde{I}$ is constructed with the Lagrangian $D$-form. 
\begin{equation}\label{tildeI}
\tilde{\mathbf{L}} = \mathbf{L}_0 - \sum_i \alpha_i(x)\big(\mathbf{L}_i - \boldsymbol{\mathrm{F}}_i\big),    
\end{equation}
where $\boldsymbol{\mathrm{F}}_i:= \mrd \boldsymbol{\mathrm{A}}_i$. The equations of motion derived from this extended action for the auxiliary fields imply that $\mrd \alpha_i(x) = 0$ and $\boldsymbol{\mathrm{F}}_i = \mathbf{L}_i$. The first condition ensures that each $\alpha_i$ is a constant over spacetime, but now as a parameter of the solution rather than the theory. The second condition, combined with the first, ensures that the on-shell dynamics of the original fields are recovered.

A crucial result of this framework is the reinterpretation of the coupling constants as conserved charges. Using the covariant phase space formalism \cite{Lee:1990gr, Wald:1993nt, Iyer:1994ys,Wald:1999wa,Ashtekar:1987hia,Ashtekar:1990gc,Barnich:2001jy,Crnkovic:1987at} (see reviews in \cite{Hajian:2015eha,Hajian:2015xlp}), it is demonstrated that each coupling $\alpha_i$ corresponds to the conserved charge associated with the (appropriately normalized) global part of the newly introduced gauge symmetry, $\boldsymbol{\mathrm{A}}_i \to \boldsymbol{\mathrm{A}}_i + \mrd\boldsymbol{\lambda}_i$ with $\mrd\boldsymbol{\lambda}_i=0$ \cite{1,Hajian:2023bhq}.

With the couplings established as conserved charges, their conjugate chemical potentials $\mathit{\Psi}^i_{_\mr{H}}$ are naturally defined as the electric potential of the corresponding gauge field $\boldsymbol{\mathrm{A}}_i$ on the black hole horizon \cite{1},
\begin{equation}\label{Psi}
\mathit{\Psi}^i_{_\mr{H}}:= \oint_{\mr{H}} \xi_{_{\mr{H}}} \cdot \mathbf{A}_i,    
\end{equation}
where $\xi_{_{\mr{H}}}$ is the horizon Killing vector. This provides a solid geometrical foundation for the chemical potentials. These new thermodynamic pairs $(\alpha_i, \mathit{\Psi}^i_{_\mr{H}})$ are then incorporated into the first law of black hole thermodynamics, leading to the extended form:
\begin{equation}
\delta M=T_{_\mr{H}}\delta S + \mathit{\Omega}_{_\mr{H}}\delta J +\mathit{\Phi}_{_\mr{H}} \delta Q+\mathit{\Psi}^i_{_\mr{H}}\delta \alpha_i.
\end{equation}
From this extended first law, a generalized Smarr formula is derived using a scaling argument \cite{Smarr:1972kt}, resulting in
\begin{equation}\label{Smarr}
(D-3) M=(D-2)T_{_\text{H}}S+(D-2)\mathit{\Omega}_{_\text{H}} J+ (D-3)\mathit{\Phi}_{_\text{H}} Q +k^{(i)}\mathit{\Psi}^i_{_\text{H}} \alpha_i,
\end{equation}
where $k^{(i)}$ is the scaling dimension of the coupling $\alpha_i$. This formulation provides a systematic and universal framework for ensuring the consistency of the Smarr formula in any covariant theory of gravity.

\section{Assessment of the universal Smarr relation}\label{Sec analysis}

This section presents examples in which the generalized Smarr formula, without the new formulation, does not hold. These cases often involve additional dimensionful parameters in the Lagrangian that are not accounted for in the standard dimensional analysis-based derivation of the Smarr formula.

\subsection{\textbf{The BTZ Black Hole in the New Massive Gravity}}
This example was previously analyzed in \cite{Hajian:2023bhq}. However, for the sake of completeness and clarity, we begin with this interesting black hole. We outline the steps in detail to assist readers who are unfamiliar with the covariant formulation of charges. Not to be pedantic, in the examples that follow, we will not lay out all the detailed calculations. The gravitational theory for this black hole is the three-dimensional New Massive Gravity (NMG), as described in \cite{Bergshoeff:2009hq}.
\begin{equation}\label{BTZ NMG}
\mathcal{L}=\frac{1}{16\pi}\left(R-2\Lambda-\beta\left(\frac{3}{8}R^2-R_{\mu\nu}R^{\mu\nu}\right)\right), \qquad \beta>0.
\end{equation}
The Banados-Teitelboim-Zanelli (BTZ) black hole, initially presented in \cite{Banados:1992wn}, is a solution of NMG as shown in \cite{Clement:2009gq}. The metric is given by:
\begin{align}
&\mrd s^2= -\Delta\mrd t^2 +\frac{\mrd r^2}{\Delta}+r^2(\mrd \varphi-\omega \mrd t)^2, \qquad \Delta:= -m+\frac{r^2}{\ell^2}+\frac{j^2}{4r^2}, \qquad \omega:= \frac{j}{2r^2}.
\end{align}
Here, the negative cosmological constant is given by $\Lambda=-\frac{1}{\ell^2}+\frac{\beta}{4\ell^4}$. The outer and inner horizons are located at $2r_\pm^2=\ell^2 (m\pm\sqrt{m^2-\frac{j^2}{\ell^2}})$, where the horizon-generating Killing vectors $\xi_\pm=\partial_t+\mathit{\Omega}_\pm \partial_\varphi$ become null, respectively. 

Although the geometry is identical to that of the BTZ black hole as a solution of the Einstein-$\ Lambda$ theory, in the NMG case,  the conserved charges are modified \cite{Clement:2009gq,Alkac:2012bz}. The mass, spin, angular velocities, temperatures, and entropies are given as 
\begin{align}\label{Prop BTZ-NMG}
& M=\left(1+\frac{\beta}{2\ell^2}\right)\frac{m}{8}, \qquad  J=\left(1+\frac{\beta}{2\ell^2}\right)\frac{j}{8},\nonumber\\
&\mathit{\Omega}_\pm=\frac{r_\mp}{\ell r_\pm}, \qquad \qquad \qquad T_\pm=\frac{r_\pm^2-r_\mp^2}{2\pi \ell^2 r_\pm}, \qquad S_\pm=\left(1+\frac{\beta}{2\ell^2}\right)\frac{\pi r_\pm}{2}.
\end{align}
Following the procedure outlined in \cite{Hajian:2023bhq}, we promote the coupling constants $\Lambda$ and $\beta$ to spacetime-dependent scalar fields $\Lambda(x)$ and $\beta(x)$. Their corresponding field strengths, ${F}_{\Lambda}(x)$ and ${F}_{\beta}(x)$—which are related to $\mathbf{F}_{\Lambda}(x)$ and $\mathbf{F}_{\beta}(x)$ in \eqref{tildeI} through a Hodge dual transformation—are then incorporated into the Lagrangian \eqref{BTZ NMG},
\begin{equation}\label{BTZ NMG tilde L}
\tilde{\mathcal{L}}=\frac{1}{16\pi}\left(R-2\Lambda(x)\left(1-{F}_{\Lambda}(x)\right)-\beta(x)\left(\frac{3}{8}R^2-R_{\mu\nu}R^{\mu\nu}-{F}_{\beta}(x)\right)\right).
\end{equation}
We have chosen normalization in defining the couplings; for instance, one may use $\frac{\Lambda}{8\pi}$ and $\frac{\beta}{16\pi}$ instead of $\Lambda$ and $\beta$ as the coupling constants, as we have chosen in \eqref{BTZ NMG tilde L}. However, such a choice of convention does not influence the physical thermodynamic laws, since these factors are automatically compensated by their inverse in the definition of the conjugate potentials.

Varying the Lagrangian with respect to the newly introduced pairs of fields, one obtains the corresponding on-shell field equations, 
\begin{equation}
{F}_{\Lambda}(x)=1, \qquad {F}_{\beta}(x)=\frac{3}{8}R^2-R_{\mu\nu}R^{\mu\nu}, \quad 
\end{equation}
along with the constancy of $\Lambda$ and $\beta$. Consequently, one has 
\begin{align}
&\mathbf{F}_{\Lambda}=\boldsymbol{\epsilon}=\sqrt{-g} \,\,\mrd t\wedge\mrd r\wedge\mrd \varphi=r \,\,\mrd t\wedge\mrd r\wedge\mrd \varphi, \\
&\mathbf{F}_{\beta}=\left(\frac{3}{8}R^2-R_{\mu\nu}R^{\mu\nu}\right)\boldsymbol{\epsilon}=\frac{3r}{2\ell^4}\,\,\mrd t\wedge\mrd r\wedge\mrd \varphi, 
\end{align}
from which one can find the gauge fields  to be 
\begin{align}
&\mathbf{A}_{\Lambda}=-\left(\frac{r^2}{2}-\frac{\beta m\ell^2}{2\beta-4\ell^2}\right)\mrd t\wedge \mrd \varphi, \label{BTZ ALam}\\
&\mathbf{A}_{\beta}=-\left(\frac{3r^2}{4\ell^4}-\frac{m(\beta-4\ell^2)}{4\ell^2(\beta-2\ell^2)}\right)\mrd t\wedge \mrd \varphi. \label{BTZ Abet}
\end{align}
Note that the second term inside each parenthesis represents a pure gauge contribution, which may be fixed through various procedures.
As an illustration, we adopt the condition that the black hole charges remain integrable (see \cite{Hajian:2021hje,Hajian:2015xlp,Ghodrati:2016vvf}).
For completeness, we explicitly demonstrate how this prescription operates within the covariant phase space formulation of conserved charges.
In the theory defined by \eqref{BTZ NMG tilde L}, and for a symmetry generator $\tilde\epsilon=\{\xi^\mu,\boldsymbol{\lambda}_\Lambda,\boldsymbol{\lambda}_\beta\}$, the variation of the corresponding charge takes the form (see Eq.~(21) in \cite{Hajian:2023bhq})
\begin{equation}\label{k modify}
\delta H_{\tilde\epsilon}=\oint_{\partial\Sigma}\tilde{\boldsymbol{k}}_{\tilde\epsilon}\,, \qquad \qquad \tilde{\boldsymbol{k}}_{\tilde \epsilon}=\boldsymbol{k}_\xi+\big(\xi\cdot\mathbf{A}_\Lambda+\boldsymbol{\lambda}_\Lambda\big)\, \delta \Lambda+\big(\xi\cdot\mathbf{A}_\beta+\boldsymbol{\lambda}_\beta\big)\, \delta \beta, 
\end{equation}
in which  $\boldsymbol{k}_\xi=\star k_\xi$ and $k^{\mu\nu}_\epsilon=k^{\mu\nu}_{0\xi }+k^{\mu\nu}_{\Lambda\xi }+k^{\mu\nu}_{\beta\xi }$, with
\begin{align}
&k_{0\xi }^{\mu\nu}=\frac{1}{16 \pi }\Big[h^{\mu\alpha}\nabla_\alpha\xi^\nu-\nabla^\mu h^{\nu\alpha}\xi_\alpha-\frac{1}{2}h \nabla^\mu\xi^\nu -(\nabla_\alpha h^{\mu\alpha}-\nabla^\mu h) \xi^\nu \Big]-[\mu\leftrightarrow\nu] \label{Einstein k}, \qquad k_{\Lambda\xi }^{\mu\nu}=0,
\end{align}
\begin{align}
&k_{\beta\xi }^{\mu\nu}=\Bigg(\frac{-3\beta}{8\times 16 \pi }\Big[\Big( h^{\mu\alpha}\nabla_\alpha\xi^\nu-\nabla^\mu h^{\nu\alpha}\xi_\alpha-\frac{1}{2}h \nabla^\mu\xi^\nu\Big)(2R)+2\Big(R^{\mu\alpha}\nabla_\alpha h-\nabla_\alpha R h^{\mu\alpha}-R^\mu_{\,\,\alpha}\nabla_\beta h^{\alpha\beta}\nonumber\\
&\hspace*{1cm}-\Box \nabla^\mu h +\nabla_\alpha\nabla^\mu\nabla_\beta h^{\alpha\beta}-\nabla^\mu(R_{\alpha\beta} h^{\alpha\beta})+\frac{1}{2}\nabla^\mu R\, h \Big)(2\xi^\nu) +\Big(R_{\alpha\beta}h^{\alpha\beta}-\nabla_\alpha\nabla_\beta h^{\alpha\beta}
+\Box h\Big)(2\nabla^\mu\xi^\nu)\nonumber\\
&\hspace*{1cm} -\Big(2R(\nabla_\alpha h^{\mu\alpha}-\nabla^\mu h)-2\nabla_\alpha R\, h^{\mu\alpha}+2\nabla^\mu R\, h\Big) \xi^\nu \Big]-[\mu\leftrightarrow\nu] \Bigg)\nonumber \\
&\hspace*{0.5cm} +\Bigg(\frac{\beta}{16 \pi }\Big[\Big(\nabla^\alpha R_\alpha^{\,\,\mu} h-\nabla_\alpha R h^{\mu\alpha}\!-\!\nabla^\mu(R_{\alpha\beta}h^{\alpha\beta})+\nabla^\mu\nabla_\alpha\nabla_\beta h^{\alpha\beta}-\nabla^\mu \Box h\Big)\xi^\nu \!+\! \Big( 2\nabla_\beta R^\mu_{\,\,\alpha}h^{\beta\nu}-2 R^{\mu\beta}\nabla_\beta h^\nu_{\,\, \alpha}\nonumber \\
&\hspace*{1cm} -2\nabla^\mu R_{\alpha\beta}h^{\nu\beta}- \nabla^\mu (\nabla_\alpha\nabla^\nu h -\nabla_\beta \nabla_\alpha h^{\nu\beta} +\Box h^\nu_{\,\,\alpha}-\nabla^\beta \nabla^\nu h_{\alpha\beta}) 
 +\nabla^\mu R^\nu_{\,\,\alpha} h +2 R^{\mu\beta} \nabla^\nu h_{\alpha\beta}\Big) \xi^\alpha \nonumber\\ 
 &\hspace*{1cm}+\Big(\nabla_\alpha \nabla^\mu h - \nabla_\beta \nabla_\alpha h^{\mu\beta}-\nabla^\beta \nabla^\mu h_{\alpha\beta}+\Box h^{\mu}_{\,\,\alpha}+2(R_{\alpha\beta}h^{\mu\beta}+R^{\mu\beta}h_{\alpha\beta})-R^{\mu}_{\,\,\alpha} h\Big) \nabla^\alpha \xi^\nu  \nonumber \\
&\hspace*{1cm} - \big(2R_{\alpha\beta}\nabla^\alpha h^{\beta \mu}-2\nabla_\alpha R^{\mu}_{\,\,\beta}h^{\alpha\beta}\!-\!R^\mu_{\,\,\alpha}\nabla^\alpha h+\nabla^\alpha R^\mu_{\,\,\alpha}h-R_{\alpha\beta}\nabla^\mu h^{\alpha\beta}\!+\!\nabla^\mu R_{\alpha\beta}h^{\alpha\beta}\big) \xi^\nu\Big]-[\mu\leftrightarrow\nu]\Bigg)\nonumber\\
&\hspace*{0.5cm} +\Bigg(\Big[ \frac{-3\delta\beta}{8\times 16 \pi }\big(4\nabla^\mu R \, \xi^\nu-2R \nabla^\mu \xi^\nu \big)+ \frac{\delta \beta}{8\pi}\big(\nabla^\mu R^\nu_{\,\,\alpha}\xi^\alpha+ R^{\nu}_{\,\,\alpha}\nabla^\alpha\xi^\mu - \nabla^\alpha R^\nu_{\,\,\alpha}\xi^\mu\big)\Big]  -[\mu\leftrightarrow\nu]\Bigg).
\end{align}
These equations require further clarification to be fully understood. 
\begin{enumerate}
\item The notation $h_{\mu\nu}=\delta g_{\mu\nu}$ as well as  $h^{\mu\nu}= g^{\mu\alpha}g^{\nu\beta}{\delta}g_{\alpha\beta}$ and $h=h^{\mu}_{\; \mu}$ is used.
\item The $k_{0\xi }^{\mu\nu}$ in \eqref{Einstein k} is the standard contribution from Einstein gravity (e.g., see Eq. (2.7) in \cite{Hajian:2016kxx}).
\item In the case of the cosmological constant $\Lambda$, the contribution to $k^{\mu\nu}_\xi$ vanishes,
\begin{equation}
k_{\Lambda\xi}^{\mu\nu}=0.    
\end{equation}
This is a general observation that holds universally and will be employed throughout the paper.
\item The tensor $k_{\beta\xi }^{\mu\nu}$ contains three distinct groups of terms.
The first two originate from the $R^2$ and $R_{\mu\nu}R^{\mu\nu}$ contributions in the Lagrangian, whose detailed derivations can be found in Sec .~3 of \cite{Ghodrati:2016vvf}.
The third group, which explicitly involves $\delta\beta$, arises from allowing the coupling $\beta$ to vary in the improved Lagrangian $\tilde{\mathcal{L}}$.
It is given by $\delta\beta$ multiplied by the associated Noether charge $Q_{\beta\xi }^{\mu\nu}$, which can be directly extracted from the first two expressions in (3.10) of \cite{Ghodrati:2016vvf}.
\item The integration in \eqref{k modify} admits any closed codimension-2 surface $\partial\Sigma$ that encloses the black hole, provided that $\tilde{\epsilon}$ generates an exact symmetry \cite{Hajian:2015xlp}—that is, the field variations vanish under the corresponding transformations.
For computational convenience, one may choose $\partial\Sigma$ to be a surface of constant time and radius.
With this choice, the integrand simplifies to
\begin{equation} \label{delH BTZ explicit}
\delta H_{\tilde\epsilon}=\int_{0}^{2\pi}\!\sqrt{-g} \, \mrd \varphi \,\tilde{{k}}^{tr}_{\tilde\epsilon},
\end{equation}
where $\tilde k^{tr}$ denotes the component $\tilde k^{\mu\nu}$ with indices $\mu=t$ and $\nu=r$. In this framework, the final expression for the charges linked to the exact symmetries, including the Killing vectors, is independent of the choice of $r$.
\item To compute the charge variations, one requires not only $\tilde \epsilon$ but also the variations of the fields. A useful choice in this context is the so-called parametric variations (see Refs. \cite{Hajian:2014twa,Hajian:2015xlp}), which correspond to variations of the fields (collectively denoted by $\Phi$) with respect to the solution parameters:
\begin{equation}
\delta \Phi=\sum_l\frac{\partial \Phi}{\partial p_l}\delta p_l,
\end{equation}
where $p_l$ labels the full set of solution parameters. In the present example, the only field is ${g_{\mu\nu},}$ and the solution parameters are $p_l \in \{m, j, \Lambda, \beta\}$.
\item From the above formulation, one can compute the variations of conserved charges, such as mass and angular momentum. When these variations are \emph{integrable}, the corresponding conserved charges can be expressed in their integrated form as functions of the parameters $p_l$.
A precise treatment of integrability is provided in the literature on the covariant formulation (see, for instance, Sec. 2 of \cite{Hajian:2015xlp}). Nevertheless, as a practical guideline within the framework of parametric variations, one may express $\delta H$ in terms of the parameters $p_l$ and check whether it takes the form $\delta \big(H(p_l)\big)$. For instance, the expression $j\delta m+ 2m \delta j$ is non-integrable, while $j\delta m+m\delta j=\delta (jm)$ is integrable.
\end{enumerate}

Having completed the detailed analysis of the charge formulation, we can now return to the evaluation of the charge and gauge fixing for the BTZ example in NMG. To compute the mass and angular momentum, the Killing generators $\tilde\epsilon_M=\{\partial_t,0,0\}$ and $\tilde\epsilon_J=\{-\partial_\varphi,0,0\}$ are employed within the framework previously outlined. The computation reveals that, for the mass and angular momentum to be integrable under the field variations,
\begin{align}
&\delta g_{\mu\nu}=\frac{\partial g_{\mu\nu}}{\partial m}\delta m +  \frac{\partial g_{\mu\nu}}{\partial j}\delta j+ \frac{\partial g_{\mu\nu}}{\partial \Lambda}\delta \Lambda + \frac{\partial g_{\mu\nu}}{\partial \beta}\delta \beta
\end{align}
the gauge freedom of $\mathbf{A}_{\Lambda}$ and $\mathbf{A}_{\beta}$ can be fixed as specified in ~\eqref{BTZ ALam} and \eqref{BTZ Abet}. Notice that the field variations $\delta\mathbf{A}_{\Lambda}$ and $\delta\mathbf{A}_{\beta}$ do not appear in the charge variation Eq. \eqref{k modify}, and so we do not need them. 
It is worth mentioning that the coupling charges $\frac{\Lambda}{8\pi}$ and $\frac{\beta}{16\pi}$ are, by construction, always integrable since they correspond to the charges of the normalized closed gauge transformations $\tilde{\epsilon}_\Lambda=\{0, \hat{\boldsymbol{\lambda}}_\Lambda,0 \}$ and $\tilde{\epsilon}_\beta=\{0, 0, \hat{\boldsymbol{\lambda}}_\beta \}$, which may be chosen as
\begin{equation}
\hat{\boldsymbol{\lambda}}_\Lambda=\frac{\mrd \varphi}{16\pi^2}, \qquad    \hat{\boldsymbol{\lambda}}_\beta=\frac{\mrd \varphi}{32\pi^2}. 
\end{equation}

With the charges and gauge field specified, substituting the fields into \eqref{Psi} yields the chemical potentials as
\begin{align}\label{Psi find}
&\mathit{\Psi}^\Lambda_\pm=\oint_\H \xi_\pm\cdot \mathbf{A}_{\Lambda}= \int_0^{2\pi}\mrd \varphi \, \xi_\pm^t\left(\frac{\beta m\ell^2}{2\beta-4\ell^2}-\frac{r^2}{2}  \right)\Bigg|_{r_\pm} = \pi\left(\frac{\beta m\ell^2}{\beta-2\ell^2}-{r_\pm^2}\right),\\ 
&\mathit{\Psi}^\beta_\pm=\!\!\oint_\H \xi_\pm\cdot \mathbf{A}_{\beta}=\!\! \int_0^{2\pi}\!\!\mrd \varphi \, \xi_\pm^t \left(\frac{m(\beta-4\ell^2)}{4\ell^2(\beta-2\ell^2)}-\frac{3r^2}{4\ell^4}  \right)\Bigg|_{r_\pm}\!\!=\pi\left(\frac{m(\beta-4\ell^2)}{2\ell^2(\beta-2\ell^2)}-\frac{3r^2_\pm}{2\ell^4}\right),
\end{align}
where $\xi_\pm^t=1$ is used. We note that the $\xi_\pm^\varphi$ term, which is $\mathit{\Omega}_\pm$, does not contribute to integration because it yields $\mathit{\Omega}_\pm \partial_\varphi\cdot\mathbf{A}_i\sim \mrd t$, whose pull-back to the horizon vanishes. For clarity, the derivation above includes the explicit steps for evaluating the integral. However, due to the similarity of subsequent cases, only the final results will be displayed for the remainder of this paper.

It can be verified that both the first law and the Smarr relation hold for each horizon:
\begin{align}
&\delta M=T_\pm \delta S_\pm+\mathit{\Omega}_\pm \delta J+\mathit{\Psi}^\Lambda_\pm \delta\left(\frac{\Lambda}{8\pi}\right)+\mathit{\Psi}^\beta_\pm \delta\left(\frac{\beta}{16\pi}\right),\\
&0=T_\pm  S_\pm+\mathit{\Omega}_\pm J-2\mathit{\Psi}^\Lambda_\pm \left(\frac{\Lambda}{8\pi}\right)+2\mathit{\Psi}^\beta_\pm \left(\frac{\beta}{16\pi}\right),
\end{align} 
respectively. The numerical coefficients $\frac{1}{8\pi}$ and $\frac{1}{16\pi}$ appearing in the coupling charges are conventional, stemming from the manner in which $\alpha_i$ and $\mathcal{L}_i$ are defined via the product $\alpha_i\mathcal{L}_i$ in the Lagrangian \eqref{tildeI}.

\subsection{\textbf{BTZ-like black hole in Horndeski gravity}}
We consider a Horndeski gravity theory in three dimensions \cite{Horndeski:1974wa} with the metric and the scalar field $\phi(x^\mu)$ as its dynamical fields, characterized by the Lagrangian:
\begin{equation}\label{Horndeski L}
\mathcal{L}=\frac{1}{16\pi}\left(R-2\Lambda-2\left(\alpha g_{\mu\nu}-\gamma G_{\mu\nu}\right)\nabla^\mu\phi\nabla^\nu\phi
\right),
\end{equation}
where $G_{\mu\nu}=R_{\mu\nu}-\frac{1}{2}Rg_{\mu\nu}$ denotes the Einstein tensor. A specific black hole solution within this theory was provided in \cite{Santos:2020xox}. The metric in the coordinates $x^\mu=(t,r,\varphi)$ is given by:
\begin{align}
& \mrd s^2=-f \mrd t^2+\frac{\mrd r^2}{f}+r^2 (\mrd\varphi-\frac{j}{r^2}\mrd t)^2, \nonumber\\
& f=-m+\frac{\alpha r^2}{\gamma}+\frac{j^2}{r^2}, \quad \mrd\phi=\sqrt{\frac{-(\alpha+\gamma\Lambda)}{2\alpha \gamma f}} \mrd r,
\end{align}
Here, $\gamma <0$, and $m,j$ represent the solution's free parameters. The thermodynamic charges and chemical potentials for this solution have been calculated, as shown in \cite{Hajian:2020dcq}:
\begin{align}
&M=\frac{(\alpha -\Lambda \gamma)m}{16 \alpha}, \qquad J= \frac{(\alpha -\Lambda \gamma)j}{8 \alpha},\qquad r_\pm^2=\frac{\gamma m \mp \sqrt{\gamma^2m^2-4\gamma\alpha j^2}}{2\alpha}, \nonumber\\
& \mathit{\Omega}_\pm=\frac{j}{r_\pm^2}, \qquad \kappa_\pm=\frac{\alpha(r^2_+-r^2_-)}{\gamma r_\pm}, \qquad T_\pm=\left(\frac{\alpha - \Lambda\gamma}{4\pi\alpha}\right) \kappa_\pm, \qquad S_\pm=\frac{\pi r_\pm}{2}.
\end{align}
For physical consistency, requiring finite and positive horizon radii necessitates $\alpha<0$. The horizon Killing vectors are $\xi_\pm=\partial_t+\mathit{\Omega}_\pm \partial_\varphi$.  One should note that the temperatures $T_\pm$ deviate from the standard Hawking temperatures $\frac{\kappa_\pm}{2\pi}$ by a specific factor. This difference is a characteristic of Horndeski gravities, where the effective graviton speed can differ from unity, as discussed in \cite{Hajian:2020dcq}. 

There are three couplings, $(\Lambda,\alpha,\gamma)$, that can be promoted to conserved charges. To this end, the Lagrangian \eqref{Horndeski L} is modified to
\begin{align}\label{Horndeski 1 tilde L}
\tilde{\mathcal{L}}=\frac{1}{16\pi}\Bigg(& R-2\Lambda(x)\left(1-{F}_{\Lambda}(x)\right)- 2\alpha(x)\left( g_{\mu\nu} \nabla^\mu\phi\nabla^\nu\phi -F_{\alpha}(x)\right)-2\gamma(x)\left( -G_{\mu\nu}\nabla^\mu\phi\nabla^\nu\phi-F_{\gamma}(x) \right)\Bigg).
\end{align}
We stress that the separation of couplings from the interaction terms by a constant factor is purely conventional; for instance, we have chosen 
$\left(\frac{\Lambda}{8\pi},\frac{\alpha}{8\pi},\frac{\gamma}{8\pi}\right)$ as the couplings in \eqref{Horndeski 1 tilde L}. Varying the Lagrangian with respect to the newly introduced field pairs yields the corresponding equations of motion, which in turn lead to the following on-shell conditions:
\begin{equation}
{F}_{\Lambda}(x)=1, \qquad {F}_{\alpha}(x)=g_{\mu\nu} \nabla^\mu\phi\nabla^\nu\phi, \qquad {F}_{\gamma}(x)=-G_{\mu\nu}\nabla^\mu\phi\nabla^\nu\phi,
\end{equation}
in parallel with the constancy of $\Lambda$, $\alpha$, and $\gamma$. Consequently, we find
\begin{align}
&\mathbf{F}_{\Lambda}=\boldsymbol{\epsilon}=\sqrt{-g} \,\,\mrd t\wedge\mrd r\wedge\mrd \varphi=r \,\,\mrd t\wedge\mrd r\wedge\mrd \varphi, \\
&\mathbf{F}_{\alpha}=\Big(g_{\mu\nu} \nabla^\mu\phi\nabla^\nu\phi\Big)\boldsymbol{\epsilon}=\frac{-(\alpha+\Lambda \gamma)r}{2\alpha \gamma}\,\,\mrd t\wedge\mrd r\wedge\mrd \varphi, \\
& \mathbf{F}_{\gamma}=\Big(-G_{\mu\nu} \nabla^\mu\phi\nabla^\nu\phi\Big)\boldsymbol{\epsilon}=\frac{(\alpha+\Lambda \gamma)r}{2 \gamma^2}\,\,\mrd t\wedge\mrd r\wedge\mrd \varphi.
\end{align}
The gauge fields associated with these field strengths $\mathbf{F}=\mrd \mathbf{A}$ are found  to be
\begin{align}
&\mathbf{A}_{\Lambda}=\left(-\frac{r^2}{2}+\frac{\gamma m}{4\alpha}\right)\mrd t\wedge \mrd \varphi, \qquad 
\mathbf{A}_{\alpha}=\left(\frac{(\alpha+\Lambda \gamma)r^2}{4\alpha \gamma}-\frac{\Lambda \gamma m}{4\alpha^2}\right)\mrd t\wedge \mrd \varphi, \nonumber \\
&\mathbf{A}_{\gamma}=\left(\frac{-(\alpha+\Lambda \gamma)r^2}{4\gamma^2}+\frac{\Lambda m}{4\alpha}\right)\mrd t\wedge \mrd \varphi.
\end{align}
The second term in each parenthesis represents the gauge-fixing term, which is determined by integrability in the same manner as described in the first example of this paper. Following the same procedure as in \eqref{Psi find}, the conjugate chemical potentials are determined as
\begin{align}
&\mathit{\Psi}^\Lambda_\pm=2\pi\left(-\frac{r_\pm^2}{2}+\frac{\gamma m}{4\alpha}\right), \qquad \mathit{\Psi}^\alpha_\pm=2\pi\left(\frac{(\alpha+\Lambda \gamma)r_\pm^2}{4\alpha \gamma}-\frac{\Lambda \gamma m}{4\alpha^2}\right), \nonumber \\
&\mathit{\Psi}^\gamma_\pm=2\pi\left(\frac{-(\alpha+\Lambda \gamma)r_\pm^2}{4\gamma^2}+\frac{\Lambda m}{4\alpha}\right).
\end{align}
The solution is characterized by five independent parameters, $m$, $j$, $\Lambda$, $\alpha$, and $\beta$ (or equivalently, by the corresponding conserved charges). The generalized first law and Smarr formula are given by:
\begin{align}
&\delta M=T_\pm \delta S_\pm+\mathit{\Omega}_\pm \delta J+\mathit{\Psi}^\Lambda_\pm \delta\left(\frac{\Lambda}{8\pi}\right)+\mathit{\Psi}^\alpha_\pm \delta\left(\frac{\alpha}{8\pi}\right)+\mathit{\Psi}^\gamma_\pm \delta\left(\frac{\gamma}{8\pi}\right),\\
&0=T_\pm  S_\pm+\mathit{\Omega}_\pm J-2\mathit{\Psi}^\Lambda_\pm \left(\frac{\Lambda}{8\pi}\right)-2\mathit{\Psi}^\alpha_\pm \left(\frac{\alpha}{8\pi}\right).
\end{align} 
These relations can be checked by varying all five solution parameters. A few remarks are worth making about the noteworthy results presented above:
\begin{enumerate}
\item This analysis involves three independent couplings, which makes the formulation highly non-trivial. The fact that it succeeds demonstrates the robustness and reliability of the conserved coupling framework.
\item The modified temperature, arising from the difference between photon and graviton speeds, is the correct quantity that consistently fits into both equations. This outcome corroborates the findings of Ref.~\cite{Hajian:2020dcq}. It is also important to emphasize that the calculation of charges and their integrability is independent of the determination of temperature. Therefore, by relying solely on this formulation and enforcing the first law, together with the Smarr relation, one can confirm the necessity of this temperature modification in the present context.
\item No contribution arises from a putative dilatonic charge. This result verifies the redundancy of the dilaton shift $\phi \to \phi+\phi_0$, in agreement with the analysis in \cite{Hajian:2016iyp}.
\item Note that $\alpha$ and $\gamma$ scale as $k^{(\alpha)}=-2$ and $k^{(\gamma)}=0$, respectively. Consequently,  $\gamma$ and its conjugates $\mathit{\Psi}^\gamma_\pm$ do not contribute to the Smarr formula in Eq. \eqref{Smarr}. 
\end{enumerate}

\subsection{\textbf{A Black Brane in Horndeski-Maxwell Gravity}}
Here, we consider a black brane solution within a Horndeski gravity theory similar to the one in Eq. \eqref{Horndeski L} with an additional Maxwell term, which is described as:
\begin{equation}\label{Horndeski L 2}
\mathcal{L}=\frac{1}{16\pi}\Big(R-F_{\mu\nu}F^{\mu\nu}-2\Lambda-2(\alpha g_{\mu\nu}-\gamma G_{\mu\nu})\nabla^\mu\phi\nabla^\nu\phi\Big).
\end{equation}
A black brane solution for this theory is characterized by the metric in coordinates $x^\mu=(t,r,x,y)$:
\begin{align}
&\mrd s^2=-h(r)\mrd t^2+\frac{\mrd r^2}{f(r)}+r^2(\mrd x^2+ \mrd y^2), \quad \mrd\phi=\sqrt{\frac{\beta-\frac{2q^2\ell^2}{3r^4}}{4\gamma f}}\mrd r, \quad A=\left(\frac{q}{r}-\frac{2q^3\ell^2}{15(4+\beta) r^5} \right) \mrd t, \nonumber\\
&h=\frac{r^2}{\ell^2}-\frac{m}{r}+\frac{4q^2}{(4+\beta)r^2}-\frac{4q^4\ell^2}{15(4+\beta)^2r^6}, \qquad f=\frac{(4+\beta)^2r^8 h}{\big(\frac{2q^2\ell^2}{3}-(4+\beta)r^4\big)^2},
\end{align}
which should satisfy the following parameter relations:
\begin{equation}
\Lambda=-\frac{3(1+\frac{\beta}{2})}{\ell^2}, \qquad \alpha=\frac{3\gamma}{\ell^2}.
\end{equation}
For this solution, the mass, electric charge, and entropy are defined as ``densities," implying they are calculated without integration over the $x$ and $y$ coordinates, as detailed in \cite{Hajian:2020dcq}. Their expressions are:
\begin{equation}
M=\frac{(4+\beta)m}{32\pi}, \qquad Q=\frac{q}{4\pi}, \qquad S=\frac{r_\H^2}{4},
\end{equation}
in which the $r_\H$ denotes the horizon radii (which are the roots of $f(r)=0$) collectively.  The surface gravity $\kappa_\H$ and electric potential $\mathit{\Phi}_{_\text{H}}$ on the horizon are:
\begin{equation}
\kappa_\H=\frac{3r_{_\text{H}}}{2\ell^2}-\frac{q^2}{(4+\beta)r^3_{_\text{H}}} , \qquad \mathit{\Phi}_{_\text{H}}=\frac{q}{r_{_\text{H}}}-\frac{2q^3\ell^2}{15(4+\beta) r^5_{_\text{H}}}.
\end{equation}
This black brane is a critical case in this study because, similar to the previous example, neither the standard nor the generalized first law nor the Smarr formula is satisfied when using the Hawking temperature $T_0=\frac{\kappa}{2\pi}$. As explained in \cite{Hajian:2020dcq}, this is a general characteristic of Horndeski gravity and other gravity models where the graviton's speed differs from unity. The physical temperature $T_\H$ in Hawking radiation, dominated by gravitons, is related to $T_0$ by a solution-dependent factor:
\begin{equation}
T_\H= \left(\frac{3(4 +\beta)r_\H ^4-2q^2\ell^2 }{24\pi r^4_{_\text{H}}}\right) \kappa_\H.
\end{equation}
Applying the coupling charge procedure to this model and its black hole solution, the improved Lagrangian is
\begin{align}\label{Horndeski 2 tilde L}
\tilde{\mathcal{L}}=\frac{1}{16\pi}\Bigg(& R-F_{\mu\nu}F^{\mu\nu}-2\Lambda(x)\left(1-{F}_{\Lambda}(x)\right)- 2\alpha(x)\left( g_{\mu\nu} \nabla^\mu\phi\nabla^\nu\phi -F_{\alpha}(x)\right)\nonumber \\
&-2\gamma(x)\left( -G_{\mu\nu}\nabla^\mu\phi\nabla^\nu\phi-F_{\gamma}(x) \right)\Bigg).
\end{align}
where the couplings $\alpha_i$ are chosen as 
$\left(\frac{\Lambda}{8\pi},\frac{\alpha}{8\pi},\frac{\gamma}{8\pi}\right)$. Similar to the previous example, we find the following on-shell conditions:
\begin{equation}
{F}_{\Lambda}(x)=1, \qquad {F}_{\alpha}(x)=g_{\mu\nu} \nabla^\mu\phi\nabla^\nu\phi, \qquad {F}_{\gamma}(x)=-G_{\mu\nu}\nabla^\mu\phi\nabla^\nu\phi,
\end{equation}
accompanied by the constancy of $\Lambda$, $\alpha$, and $\gamma$. Consequently, we find
\begin{align}
&\mathbf{F}_{\Lambda}=\boldsymbol{\epsilon}=\sqrt{-g} \,\,\mrd t\wedge\mrd r\wedge\mrd x \wedge \mrd y=\left(\frac{12r^4 - 2\ell^2q^2 + 3\beta r^4}{3r^2(\beta + 4)}\right) \,\,\mrd t\wedge\mrd r\wedge\mrd x \wedge \mrd y, \\
&\mathbf{F}_{\alpha}=\Big(g_{\mu\nu} \nabla^\mu\phi\nabla^\nu\phi\Big)\boldsymbol{\epsilon}=\Big(\frac{3 \beta  r^4-2 \ell^2 q^2}{12 \gamma r^4}\Big)\left(\frac{12r^4 - 2\ell^2q^2 + 3\beta r^4}{3r^2(\beta + 4)}\right)\,\,\mrd t\wedge\mrd r\wedge\mrd x \wedge \mrd y, \\
& \mathbf{F}_{\gamma}=\Big(-G_{\mu\nu} \nabla^\mu\phi\nabla^\nu\phi\Big)\boldsymbol{\epsilon}=\Big(\frac{ 3\beta r^{4}-2\ell^{2}q^{2}}{4\ell^{2}\gamma r^{4}}\Big)\left(\frac{12r^4 - 2\ell^2q^2 + 3\beta r^4}{3r^2(\beta + 4)}\right)\,\,\mrd t\wedge\mrd r\wedge\mrd x \wedge \mrd y.
\end{align}
The gauge fields associated with these field strengths $\mathbf{F}=\mrd \mathbf{A}$ are found  to be
\begin{align}
&\mathbf{A}_{\Lambda}=-\left(\frac{r^3}{3} + \frac{2 \ell^2 q^2}{(3 \beta  + 12)r}\right)\mrd t\wedge \mrd x \wedge \mrd y, \nonumber \\
& \mathbf{A}_{\alpha}=-\left(\frac{\beta r^{3}}{12}
+ \frac{\ell^{2} q^{2}(\beta+2)}{3\gamma(\beta+4)}\frac{1}{r}
- \frac{\ell^{4} q^{4}}{45\gamma(\beta+4)}\frac{1}{r^{5}}
\right)\mrd t\wedge \mrd x \wedge \mrd y, \qquad \mathbf{A}_{\gamma}=\frac{3}{\ell^2}\mathbf{A}_{\alpha}.
\end{align}
The solution is characterized by five independent parameters: $(m,q,\beta,\ell,\gamma)$, where $\beta$ and $\ell$ effectively represent $\Lambda$ and $\alpha$, respectively. Following similar steps as in the previous examples, the conjugate chemical potentials are found as
\begin{align}
&\mathit{\Psi}^\Lambda_\H=-\left(\frac{r_\H^3}{3} + \frac{2 \ell^2 q^2}{(3 \beta + 12)r_\H}\right), \nonumber\\
& \mathit{\Psi}^\alpha_\H=-\left(\frac{\beta r_\H^{3}}{12}
+ \frac{\ell^{2} q^{2}(\beta+2)}{3\gamma(\beta+4)}\frac{1}{r_\H}
- \frac{\ell^{4} q^{4}}{45\gamma(\beta+4)}\frac{1}{r_\H^{5}}
\right), \qquad\mathit{\Psi}^\gamma_\H=\frac{3}{\ell^2}\mathit{\Psi}^\alpha_\H.
\end{align}
The generalized first law and Smarr formula for this example are given by:
\begin{align}
&\delta M=T_\H \delta S_\H+\mathit{\Phi}_{_\text{H}}\delta Q+\mathit{\Psi}^\Lambda_\H \delta\left(\frac{\Lambda}{8\pi}\right)+\mathit{\Psi}^\alpha_\H \delta\left(\frac{\alpha}{8\pi}\right)+\mathit{\Psi}^\gamma_\H \delta\left(\frac{\gamma}{8\pi}\right),\\
&M=2T_\H  S_\H+\mathit{\Phi}_{_\text{H}} Q-2\mathit{\Psi}^\Lambda_\H \left(\frac{\Lambda}{8\pi}\right)-2\mathit{\Psi}^\alpha_\H \left(\frac{\alpha}{8\pi}\right),
\end{align} 
which can be checked by varying all five solution parameters $(m,q,\Lambda,\alpha,\gamma)$. A few comments can be illuminating regarding the analysis conducted above. 
\begin{enumerate}
\item Similar to the previous example, we note that $\gamma$ is dimensionless; hence, $k^{(\gamma)} = 0$ is present in the universal Smarr relation, Eq. \eqref{Smarr}. Consequently, $\gamma$ and its conjugate quantity $\mathit{\Psi}^\gamma_\H$ do not contribute to the above relation.

\item We observe that modifying the Hawking temperature within Horndeski gravity \cite{Hajian:2020dcq} offers an effective framework for describing the thermodynamic behavior of its black hole solutions.
 An exact gauge symmetry must be accompanied.
\item In gravitational theories coupled with gauge fields, the Killing vectors should be concomitant with an accurately adjusted gauge transformation to reproduce charge variations within the framework of parametric variations and the covariant phase space (jointly referred to as the “Solution Phase Space Method” \cite{Hajian:2015xlp}). Specifically, the appropriate entropy generator is the horizon Killing vector $\xi_\H$ together with the gauge transformation $\lambda_\H = -\frac{2\pi}{\kappa_\H}\mathit{\Phi}_{_\text{H}}$. Fundamentally, this modification ensures both the integrability condition \cite{Hajian:2015xlp} and the gauge independence of the first law \cite{Hajian:2022lgy}. 
\item The normalization of the entropy generator becomes ill-defined in the extremal limit, as $\kappa_\H$ vanishes. Nonetheless, one can define an infinite family of entropy generators in the near-horizon region of the extremal geometry. For further details, the interested reader is referred to the series of works \cite{Hajian:2014twa,Hajian:2013lna,Hajian:2017mrf,Compere:2015bca,Compere:2015mza}, reviewed in \cite{Hajian:2015eha}, which examine these generators and their implications for the laws of black hole thermodynamics.
 
\end{enumerate}

\subsection{\textbf{Martinez-Teitelboim-Zanelli (MTZ) Black Hole}}

As our next black hole, we focus on a theory that involves the metric $g_{\mu\nu}$, a scalar field $\phi$, and the Maxwell gauge field $A_\mu$ as its dynamical components \cite{Martinez:2002ru,Barlow:2005yd}. The Lagrangian is:
\begin{equation}
\mathcal{L}=\frac{1}{16\pi }\left(R-2\Lambda-F_{\mu\nu}F^{\mu\nu}-2\nabla_\mu \phi \nabla^\mu \phi -\frac{1}{3}R\phi^2-\alpha \phi^4\right).
\end{equation}
A static and spherically symmetric black hole solution of this theory is given as:
\begin{align}\label{MTZ}
&\mrd s^2= -f\mrd t^2 +\frac{\mrd r^2}{f}+r^2 (\mrd \theta^2 + \sin^2\theta \mrd \varphi^2), \qquad f=(1-\frac{m}{r})^2-\frac{r^2}{\ell^2},\nonumber\\
&A=\frac{q}{r}\mrd t, \qquad \phi=\frac{\sqrt{3(m^2-q^2)}}{r-m},
\end{align}
whose parameters satisfy:
\begin{equation}\label{MTZ q relation}
\Lambda=\frac{3}{\ell^2}, \, \qquad q^2=m^2(1+\frac{2\Lambda}{9\alpha}).
\end{equation}
The black hole horizon radii are $r_\pm=\frac{\ell}{2}(\pm 1\mp\sqrt{1\mp\frac{4m}{\ell}})$, and a cosmological horizon exists at $r_c=\frac{\ell}{2}(1+\sqrt{1-\frac{4m}{\ell}})$. For black hole existence, the conditions $0 < 4m < \ell$ and $\frac{2\Lambda}{9}<|\alpha|$ must hold. The thermodynamic properties of this black hole, known as the four-dimensional MTZ black hole, including mass, electric charge, horizon potential, temperature, and entropy, have been determined and are presented in \cite{Martinez:2002ru,Winstanley:2004ay,Barlow:2005yd}:
\begin{align}
&M=m,\quad Q=q, \quad \mathit{\Phi}_\H=\frac{q}{r_\H}, \quad T_\H=\frac{m(r_\H-m)}{2\pi r_\H^3}-\frac{\Lambda r_\H}{6\pi}, \quad S_\H=\pi r_\H^2\left(1-\frac{m^2-q^2}{(r_\H-m)^2}\right).
\end{align}
In the relations above, the temperature $T_\H$ corresponds to the standard Hawking temperature, calculated as $T_\H=\frac{1}{4\pi}\frac{df}{dr}$ at the horizon. The entropy $S_\H$ is equivalent to the Bekenstein-Hawking entropy $\frac{A_\H}{4}$, scaled by a factor related to the scalar curvature in the Lagrangian.

We observe that the electric charge in this solution is not an independent parameter; instead, it is determined by the couplings $(\Lambda,\alpha)$ and the solution parameter $m$ through the relation \eqref{MTZ q relation}. The following analysis treats the coupling constants $(\Lambda,\alpha)$ as solution parameters, so that the thermodynamics of the MTZ black hole can be described in terms of three solution parameters $(m, \Lambda,\alpha)$, corresponding to the three associated conserved charges.

By applying the coupling charge procedure to this model and its black hole solution, the modified Lagrangian takes the form
\begin{align}\label{tilde Horndeski 2 tilde L}
\tilde{\mathcal{L}}=\frac{1}{16\pi}\Bigg( R-F_{\mu\nu}F^{\mu\nu}-2\nabla_\mu\phi\nabla^\mu\phi-\frac{1}{3}R\phi^2-2\Lambda(x)\left(1-{F}_{\Lambda}(x)\right) 
-\alpha(x)\left( \phi^4- {F}_{\alpha}(x) \right)\Bigg),
\end{align}
where the couplings $\alpha_i$ in Eq.\eqref{tildeI} are selected as
$\left(\frac{\Lambda}{8\pi},\frac{\alpha}{16\pi}\right)$. From the field equations, we obtain the following on-shell conditions:
\begin{equation}
{F}_{\Lambda}(x)=1, \qquad {F}_{\alpha}(x)=\phi^4,
\end{equation}
together with the constancy of $\Lambda$ and $\alpha$. The solution is therefore characterized by three independent parameters: $(m,\Lambda,\alpha)$.  As a result of Hodge duality, we have
\begin{align}
&\mathbf{F}_{\Lambda}=\boldsymbol{\epsilon}=\sqrt{-g} \,\,\mrd t\wedge\mrd r\wedge \mrd \theta \wedge\mrd \varphi=r^2\sin \theta \,\,\mrd t\wedge\mrd r \wedge \mrd \theta \wedge\mrd \varphi, \\
&\mathbf{F}_{\alpha}=\phi^4\boldsymbol{\epsilon}=\left(\frac{{9(m^2-q^2)^2}}{(r-m)^4}\right)r^2\sin \theta\,\,\mrd t\wedge\mrd r\wedge \mrd \theta \wedge \mrd \varphi \label{F alpha MTZ}.
\end{align}
The gauge potentials corresponding to these field strengths $\mathbf{F}=\mrd \mathbf{A}$ are found to be
\begin{align}
\mathbf{A}_{\Lambda}&=-\frac{r^3}{3} \sin \theta \, \mrd t\wedge \mrd \theta \wedge \mrd \varphi,  \\
\mathbf{A}_{\alpha}&=\left(\frac{3(m^2-q^2)^2(m^2-3mr+3r^2)}{(r-m)^3}+\frac{3(m^2-q^2)^2}{m}\right)\sin \theta\, \mrd t\wedge \mrd \theta \wedge \mrd \varphi \label{A alpha MTZ}\\
&=\frac{3r^3(m^2-q^2)^2}{m(r-m)^3}\sin \theta\, \mrd t\wedge \mrd \theta \wedge \mrd \varphi \nonumber 
\end{align}
The calculation in Eq. \eqref{A alpha MTZ} is presented explicitly to highlight that the first term within the parentheses arises from the integration of the field strength \eqref{F alpha MTZ}, whereas the second term, which is the integration constant, corresponds to the gauge-fixing contribution stemming from the integrability of the covariant mass in the parameter space defined by $m$ and $\alpha$.

Using the definition of the coupling conjugate potentials in Eq. \eqref{Psi} and adopting the horizon Killing vector $\xi_\H=\partial_t$, we proceed as in the previous examples to obtain
\begin{align}
\mathit{\Psi}^\Lambda_\H=-\frac{4\pi r_\H^3}{3}, \qquad \mathit{\Psi}^\alpha_\H=\frac{12\pi r_\H^3(m^2-q^2)^2}{m(r_\H-m)^3}.
\end{align}
Accordingly, the generalized first law and Smarr relation for this case read:
\begin{align}
&\delta M=T_\H \delta S_\H+\mathit{\Phi}_{_\text{H}} \delta Q+\mathit{\Psi}^\Lambda_\H \delta\left(\frac{\Lambda}{8\pi}\right)+\mathit{\Psi}^\alpha_\H \delta\left(\frac{\alpha}{16\pi}\right), \\
&M=2T_\H S_\H+\mathit{\Phi}_{_\text{H}} Q-2\mathit{\Psi}^\Lambda_\H \left(\frac{\Lambda}{8\pi}\right)-2\mathit{\Psi}^\alpha_\H \left(\frac{\alpha}{16\pi}\right),
\end{align}
which can be verified by directly substituting the thermodynamic quantities and varying the three solution parameters $(m,\Lambda,\alpha)$.

The analysis of this example shows that both the generalized first law and the universal Smarr relation hold for the cosmological horizon and for black hole horizons. Moreover, we find that the electric charge $Q$
contributes to these relations, even though it is not an independent parameter in this particular black hole solution.

\subsection{\textbf{AdS-Schwarzschild in Higher Curvature Gravity}}

Here, we investigate the theory of Einstein-$\Lambda$ gravity augmented with higher curvature terms, applicable in arbitrary dimensions $D\geq 3$. The Lagrangian is expressed as:
\begin{equation}
\mathcal{L}=\frac{1}{16\pi}\Big(R-2\Lambda+\alpha R^2+\beta R_{\mu\nu}R^{\mu\nu}\Big),
\end{equation}
where $\alpha$ and $\beta$ are arbitrary constants.
A solution to this theory is the generalization of the AdS-Schwarzschild black hole to $D$ dimensions:
\begin{align}
\mrd s^2=-f\mrd t^2+\frac{\mrd r^2}{f}+r^2 \mrd \Omega_{_{D-2}}^2, \qquad  f=1-\frac{2m}{r^{D-3}}+\frac{r^2}{\ell^2}, \\ 
\Lambda =\frac{-\ell^2(D^2-3D+2)+(\alpha D + \beta)(D-4)(D-1)^2}{2\ell^4}. \label{Schd Lambda}
\end{align}
The conserved charges, including mass and entropy, for these black holes are influenced by both the solution and the underlying theory. While these solutions resemble standard AdS-Schwarzschild black holes, the inclusion of higher curvature terms distinguishes the theory from basic Einstein-$\Lambda$ gravity. The modified charges, as determined in \cite{Ghodrati:2016vvf}, are:
\begin{align}
&M= \frac{(D-2)\Omega_{_{D-2}}}{8\pi} m \mathcal{X},\qquad T_\H=\frac{(D-1)r_{_\text{H}}^{D-2}+(D-3)\ell^2\,r_{_\text{H}}^{D-4}}{4\pi\ell^2\,r_{_\text{H}}^{D-3}}, \qquad S_{_\text{H}}= \frac{r_{_\text{H}}^{D-2}\Omega_{_{D-2}}}{4} \mathcal{X}
\end{align}
where:
\begin{equation}\label{Prop. AdS-Schw-higher}
\mathcal{X}=\frac{\ell^2- 2D(D-1)\alpha- 2(D-1) \beta}{\ell^2}\,, \qquad \Omega_{_{D-2}}=\frac{2\pi^{\frac{D-1}{2}}}{\Gamma(\frac{D-1}{2})}\,,
\end{equation}
and the horizons are defined by $r_{_\text{H}}^{D-1}+\ell^2 r_{_\text{H}}^{D-3}-2m\ell^2=0$. Besides, in these geometries, the Ricci tensor and scalar are
\begin{align}\label{Schd Ricci}
R_{\mu\nu}=-\frac{D-1}{\ell^2}g_{\mu\nu}, \qquad R=-\frac{D(D-1)}{\ell^2}.
\end{align}

There are four constants in this configuration: one solution parameter, $m$, and three couplings, $(\Lambda, \alpha, \beta)$, subject to the constraint \eqref{Schd Lambda}. To treat all four as solution parameters, the Lagrangian can be extended as  
\begin{equation}\label{Schd tilde L}
\tilde{\mathcal{L}}=\frac{1}{16\pi}\left(R-2\Lambda(x)\left(1-{F}_{\Lambda}(x)\right)-\alpha(x)\left(-R^2-{F}_{\alpha}(x)\right)-\beta(x)\left(-R_{\mu\nu}R^{\mu\nu}-{F}_{\beta}(x)\right)\right),
\end{equation}
where the couplings $\alpha_i$ appearing in Eq.\eqref{tildeI} are chosen as  
$\left(\frac{\Lambda}{8\pi},\frac{\alpha}{16\pi},\frac{\beta}{16\pi}\right)$. From the field equations, the following on-shell relations are obtained:  
\begin{equation}
{F}_{\Lambda}(x)=1, \qquad {F}_{\alpha}(x)=-R^2, \qquad {F}_{\beta}(x)=-R_{\mu\nu}R^{\mu\nu},
\end{equation}
accompanied by the constancy of $\Lambda$, $\alpha$, and $\beta$. Thus, the solution is characterized by four (not completely independent) parameters, $(m,\Lambda,\alpha,\beta)$. By invoking Hodge duality and using Eqs. \eqref{Schd Ricci}, we further find  
\begin{align}
&\mathbf{F}_{\Lambda}=\boldsymbol{\epsilon}=r^{D-2}\,\,\mrd t\wedge\mrd r\wedge\mrd \Omega_{D-2}, \\
&\mathbf{F}_{\alpha}=-R^2\boldsymbol{\epsilon}=-\frac{D^2(D-1)^2}{\ell^4}r^{D-2}\,\,\mrd t\wedge\mrd r\wedge\mrd \Omega_{D-2}, \label{F alpha Schd}\\
&\mathbf{F}_{\beta}=-R_{\mu\nu}R^{\mu\nu}\boldsymbol{\epsilon}=-\frac{D(D-1)^2}{\ell^4}r^{D-2}\,\,\mrd t\wedge\mrd r\wedge\mrd \Omega_{D-2}. \label{F beta Schd}
\end{align}
By integrating over the radial coordinate, the corresponding gauge potentials are obtained as  
\begin{align}
&\mathbf{A}_{\Lambda}=\left(-\frac{r^{D-1}}{D-1}+m\frac{\mrd \mathcal{X}}{\mrd \Lambda} \right)  \,\mrd t\wedge \mrd \Omega_{D-2}, \nonumber \\ 
&\mathbf{A}_{\alpha}= \left(\frac{D^2(D-1)r^{D-1}}{\ell^4}+2m\frac{\mrd \mathcal{X}}{\mrd \alpha}\right) \,\mrd t\wedge \mrd \Omega_{D-2},\nonumber\\ 
&\mathbf{A}_{\beta}= \left(\frac{D(D-1)r^{D-1}}{\ell^4}+2m\frac{\mrd \mathcal{X}}{\mrd \beta}\right)\,\mrd t\wedge \mrd \Omega_{D-2}. \label{A Schd}
\end{align}
The derivatives of  $\mathcal{X}$ which appear as the gauge fixing terms are basically derivatives in the parameter space $(\Lambda, \alpha, \beta)$ ($\ell$ can be used instead of $\Lambda$ via the chain rule). 

By employing the definition of coupling conjugate potentials from Eq.~\eqref{Psi} and selecting the horizon Killing vector $\xi_\H=\partial_t$, we replicate the procedure from the previous cases to obtain 
\begin{align}
&\mathit{\Psi}^\Lambda_\H=\left(-\frac{r_\H^{D-1}}{D-1}+m\frac{\mrd \mathcal{X}}{\mrd \Lambda} \right)\Omega_{D-2}, \nonumber \\
&\mathit{\Psi}^\alpha_\H=\left(\frac{D^2(D-1)r_\H^{D-1}}{\ell^4}+2m\frac{\mrd \mathcal{X}}{\mrd \alpha}\right)\Omega_{D-2}, \nonumber \\
&\mathit{\Psi}^\beta_\H= \left(\frac{D(D-1)r_\H^{D-1}}{\ell^4}+2m\frac{\mrd \mathcal{X}}{\mrd \beta}\right)\Omega_{D-2}.
\end{align}
The generalized first law and Smarr formula for this example are given by:
\begin{align}
&\delta M=T_\H \delta S_\H+\mathit{\Psi}^\Lambda_\H \delta\left(\frac{\Lambda}{8\pi}\right)+\mathit{\Psi}^\alpha_\H \delta\left(\frac{\alpha}{16\pi}\right)+\mathit{\Psi}^\beta_\H \delta\left(\frac{\beta}{16\pi}\right),\\
&(D-3)M=(D-2)T_\H  S_\H-2\mathit{\Psi}^\Lambda_\H \left(\frac{\Lambda}{8\pi}\right)+2\mathit{\Psi}^\alpha_\H \left(\frac{\alpha}{16\pi}\right)+2\mathit{\Psi}^\beta_\H \left(\frac{\beta}{16\pi}\right).
\end{align}

\section{Conclusions}
The analysis presented in the paper successfully demonstrates the simplicity and reliability of the recent {universal theoretical framework for black hole thermodynamics}, which consistently incorporates all dimensionful coupling constants into the first law and the celebrated Smarr formula. The application of this conserved coupling framework across several complex black hole solutions confirms its robustness and universality:

\begin{enumerate}
    \item[-] New Massive Gravity (NMG) and Horndeski Gravity: The framework successfully reconciles the thermodynamic relations for the BTZ black hole in NMG, incorporating two couplings ($\Lambda$ and $\beta$). The analysis of the BTZ-like black hole in Horndeski gravity, involving three independent couplings ($\Lambda, \alpha, \gamma$), further demonstrates the formulation's reliability.
    \item[-] Modified Temperature Consistency: In the Horndeski gravity examples (BTZ-like black hole and Black Brane), the formulation confirmed that the modified temperature (arising from the difference between photon and graviton speeds) is the correct physical quantity that consistently satisfies both the extended first law and the generalized Smarr relation. This outcome corroborates previous findings and underscores the need for appropriate temperature adjustments and precise gauge symmetry.
    \item[-] Dimensionless Couplings: The analysis showed that dimensionless couplings, such as $\gamma$ in the Horndeski examples, do not contribute to the Smarr formula, consistent with the theoretical scaling argument $k^{(i)}=0$.
    \item[-] Higher-Dimensional Examples: The MTZ black hole analysis confirmed the validity of the generalized first law and Smarr relation, even though the electric charge $Q$ was not an independent parameter but was determined by the couplings and solution parameter $m$. Likewise, the AdS–Schwarzschild solution in higher-curvature gravity for arbitrary dimensions $D \ge 3$ confirmed that the extended thermodynamic laws remain consistent, even in the presence of constraints among the couplings.
\end{enumerate}

The results obtained through this approach demonstrate the necessity of a fully generalized formulation that consistently includes all dimensionful parameters. By properly enlarging the theory, this framework confirms that the Smarr formula is not specific to General Relativity, but rather a {universal integrated relation} among thermodynamic coordinates and their conjugates.

\vskip 1 cm

\noindent \textbf{Acknowledgments:} K.H. thanks M. H. Vahidinia for his helpful comments. This work has been supported by the TUBITAK International Researchers Program No. 2221.

\end{document}